# Light-controllable chiral dopant based on azo-fragment: synthesis and characterization

V. Chornous[a], V. Bratenko[a], M. Vovk[b], Yu. Dmytriv[c,d], A. Rudnichenko[c,e], M. Skorobagatko[f], N. Kasian[g], L. Lisetski[g] and I. Gvozdovskyy[h]*

[a]*Department of Medical and Pharmaceutical Chemistry, Bukovinian State Medical University, Chernivtsi, Ukraine;* [b]*Department of Mechanisms of Organic Reactions, Institute of Organic Chemistry of the National Academy of Sciences of Ukraine, Kyiv, Ukraine;* [c]*Enamine Ltd, Kyiv, Ukraine,* [d]*Department of Chemical Technology, National Technical University of Ukraine "Igor Sikorsky Kyiv Polytechnic Institute", Kyiv, Ukraine;* [e] *ChemBioCenter, National Taras Shevchenko University, Kyiv, Ukraine;* [f]*Department of Optical and Optoelectronic Instruments, National Technical University of Ukraine "Igor Sikorsky Kyiv Polytechnic Institute", Kyiv, Ukraine;* [g]*Department of Molecular and Heterostructured Materials, Institute for Scintillation Materials of STC "Institute for Single Crystals" of the National Academy of Sciences of Ukraine, Kharkiv, Ukraine;* [h]*Department of Optical Quantum Electronics, Institute of Physics of the National Academy of Sciences of Ukraine.*

Institute of Physics of the National Academy of Sciences of Ukraine, Prospekt Nauki 46, Kyiv-28, 03028, Ukraine, telephone number: +380 44 5250862, *E-mail: igvozd@gmail.com

# Light-controllable chiral dopant based on azo-fragment: synthesis and characterization

We present the newly synthesized chiral dopant 2-[(2-isopropyl-5-methylcyclohexyl)oxy]-2-oxoethyl 4-{(E)-[4-(decyloxy)phenyl]diazenyl}benzoate (ChD-3501), consisting of azo- and aliphatic fragments together with a chiral center based on *l*-menthol as a reversible light-controllable chiral dopant. To assess the effects of UV/VIS irradiation and temperature in the isotropic and liquid crystalline (LC) states, we studied the spectral kinetics of ethanol solution of ChD-3501, as well as induction of the cholesteric helix when it was dissolved in nematic LC (E7) as a chiral dopant. The concentration dependence of the helical pitch of the induced cholesterics was studied by means on Grandjean-Cano method, and the helical twisting power of ChD-3501 in the nematic host E7 was determined. The reversible *trans-cis* isomerization of chiral dopandt ChD-3501 in E7 under UV/VIS irradiation was studied, and it has been found that the storage of the *cis*-isomer at certain constant temperature also leads to the reversible isomerisation, which presents a certain interest for applications.

1.     **Introduction**

Liquid crystals as a new state of condensed matter were discovered about 160 years ago by Planer [1,2] and Reinitzer (in collaboration with Lehmann) [3-5] in the course of their investigation of cholesterol esters. However, the interest in studying cholesteric liquid crystals (CLCs) is still very high, in particular, in the light-sensitive CLCs. The recent studies concentrated on chiral dopants (ChDs) possessing various characteristics as high chirality, [6-15] selective light absorption [16-22] *etc.*, and on their various applications as liquid crystalline optical elements with light-controllable characteristics. [17,23-37]

It is known that, alongside cholesterol esters, the cholesteric properties can be exhibited by nematic-like compounds with asymmetric side chains (so-called chiral nematic) or mixtures based on nematic liquid crystals (NLCs) and optically active (chiral) compounds, which may be mesogenic or nonmesogenic (so-called induced CLCs). [38,39]

The dissolution of small amounts of ChDs in NLCs results in the formation of the helical mesophase characterized by the preferred orientation of the long axes of the anisometric molecules (director $\vec{n}$), rotating uniformly along the helix axis.[38-41] The reason for the appearance of induced CLCs is the ability of the ChD molecules with a concentration $C$ to impose long-range twist upon NLCs by forming the helical structure possessing a certain helical pitch $P$. This ability is based on the helical-twisting power (HTP, $\beta$) of the ChD molecules and can be expressed as a $\beta = 2 \times (P \times C)^{-1}$. [41,42] In addition, the twisting sense of cholesteric helix can be right- or left-handed, depending upon the nature of ChDs and the molecular interaction between ChDs and NLCs. [38-42] Under certain conditions (*i.e.* high concentration $C$ of ChDs or high value of HTP) the helical structure can selectively reflect light. A phenomenon of selective reflection of light (*i.e.* Bragg diffraction at the maximum wavelength $\lambda_{max} = <n> \times P$, where $<n>$ is the average refractivity index) is most clearly observed for the *planar* texture of the CLCs, when the helix axis and the direction of observation coincide. If the helical axis is perpendicular to the observation plane, the *homeotropic* texture of CLCs (or the so-called "*fingerprint*" texture) can be observed. [5,41] In the recent years, the usage of planar/homeotropic textures is often proposed for various practical applications in optics. [23,26,27,29-31,33,36-39,42-44]

It is known that cholesteric helix can be sensitive to various external factors (*e.g.* temperature, [45-47] mechanical stress, [46,48,49] chemical interaction, [50-53] electric

or magnetic fields [44,54-63] and light [14,20,21,30,36,64-67]). As shown by groups of B. Feringa [14,20,68-75] and Q. Li [16,21,76-79] the usage of light as an instrument for controlling the chirality of the molecules can be interesting for the practical application in LC as light sensitive chiral switches (or so-called chiral molecular motors). The light-induced isomerisation of ChD molecules could lead to changes in HTP, and consequently, in the maximum wavelength of reflected light $\lambda_{max}$ of the CLCs. The dynamic changes of isomers and parameters of the self-organized helical structure provide the fertile ground for both the understanding of the nature of chirality from the point of view of fundamental science [38,39,80,81] and the various applications of light-controllable helical structures [16,21,22,29-31,33,36,37,52,64,67,68,71,73,82,83]. Such helical structures have been applied in tunable colour filters and LC lasers [26,29,31,34,35,77,84], optically flexible displays [10,85], diffraction gratings working in regime of Raman-Nath [30,36,64,86] and also for the switching of blue phases liquid crystals (BPLCs) [87-92].

It is known that azobenzene fragment is often used for synthesis of compounds possessing the reversible light-controllable isomerisation to create various chiral dopants [13-17,19,77-79] and chiral polymer layers [72,93,94] in order to induce the helical structure with certain properties (*i.e.* high HTP, photoferroelectricity [95]).

In this manuscript we will describe synthesis of a novel chiral dopant that comprises a light-sensitive moiety based on azobenzene and aliphatic fragments with *l*-menthyl chiral centre ensuring efficient helical twisting and carry out a rather comprehensive characterization of the properties of the induced cholesteric mixtures. As an example of possible applications, we consider the light-controllable cholesteric diffraction grating of the Raman-Nath type.

## 2. Experiment

For the synthesis of the 2-[(2-isopropyl-5-methylcyclohexyl)oxy]-2-oxoethyl 4-{(E)-[4-(decyloxy)phenyl]diazenyl}benzoate (further denoted as ChD-3501), the initial materials and reagents from Enamine Ltd (Kyiv, Ukraine) were used; the "p.a." grade solvents were used.

IR spectra were recorded using a Bruker Vertex 70 spectrometer in KBr pellets. [1]HNMR spectra were acquired on a Varian VXR-400 (400 MHz) spectrometer in DMSO-$d_6$ with TMS as internal standard. Chemical shifts ($\delta$) and J values are given are given in ppm and Hz, respectively. Melting points were determined by a Kofler bench and are uncorrected.

To study of the influence of UV/VIS radiation and temperature on the reversible *trans-cis* isomerisation, the ethanol solution of ChD-3501, possessing initial optical density OD ~ 2 in quartz cuvette with thickness 1 cm, was prepared. The ethanol solution was of light orange colour.

As a chiral compound to induce the cholesteric helix, the synthesized ChD-3501 in concentrations of 0.2 – 8.5 wt.%, was used. The nematic LC E7 (a mixture of cyano-biphenyl and terphenyl molecules) was obtained from Licrystal, Merck (Darmstadt, Germany). [97] The nematic E7 has a positive dielectric anisotropy $\Delta\varepsilon = +13.8$ and the nematic-isotropic (N-Iso) transition temperature $T_{Iso} = 58$ °C. [97]

To obtain the *planar* (Grandjean-Cano texture) alignment of the induced CLC, n-methyl-2-pyrrolidone solution of the polyimide PI2555 (HD MicroSystems, USA) in proportion 10:1 was used. PI2555 solution was spin-coated (6800 rpm, 10 s) on glass substrates and further annealed for 30 min at a temperature of 180 degrees of Celsius. Thereafter, thin films of PI2555 were rubbed in one direction several times the number of rubbings $N_{rubb}$ = 10 corresponds to strong azimuthal anchoring energy $W_\varphi = (4 \pm 1) \times 10^{-4}$ J/m$^2$. [42,43]

To assemble the wedge-like LC cells, two substrates with opposite rubbing directions on both aligning PI2555 films were used. The thickness of the thick end $d$ of the wedge-like LC cell was set to about 25 - 30 µm by Mylar spacers.

To measure the gap between the thin end $d_0$ and thick end $d$ of the wedge-like LC cell, the interference method measuring the transmission spectrum of empty cell by means of spectrometer (Ocean Optics 4000USB, USA) was used. The wedge-like LC cells were filled with induced CLCs using capillary action at the temperatures of the isotropic phase and slowly cooled to the room temperature at a rate of ~ 0.2 ºC/min.

To determine of the length of cholesteric pitch $P$, the Grandjean-Cano method was used. [97,98] It is known that in the wedge-like LC cell, with thickness of thin end $d_0$ and thickness of thick end $d$, the cholesteric helix pitch $P$ is related to the number $N_C$ of the Grandjean-Cano bands, which differ from each other by half-pitch $P/2$, as follows:

$$P = \frac{2 \times (d - d_0)}{N_c} \qquad (1)$$

To find out the twisting sense of the induced cholesteric helix, we used a recently proposed rapid method, [99-101] which is based on determination of the rotation direction of a small single crystal of chiral compound dissolving at the top of nematic droplet. In addition, the twisting sense of the cholesteric helix induced by ChD-3501 in nematic host E7 was also shown within the Grandjean-Cano method by the colour shift in the wedge-like LC cell. [102] For the right-handed (*i.e.*, wave number $q_0 = 4\pi \times \beta \times C$ is positive, $q_0 > 0$) cholesteric helix, the interference fringes are shifted towards regions at the thin end of wedge when the polarizer (analyzer) rotates in the clockwise (or counterclockwise, depending upon observation conditions) direction, and vice versa for the left-handed ($q_0 < 0$) cholesteric helix. [102]

The Grandjean-Cano texture of the CLC was observed using crossed polarizers, and images were taken by digital camera (Nikon D80, Japan). Additionally, to estimate the helical screw sense of the chiral mixtures for various UV exposures, we used the Grandjean-Cano method as described previously. [38,39,102]

Illumination of the ethanol solution of the ChD-3501 in the 1 cm quartz cuvette and of the wedge-like LC cells containing the cholesteric mixtures with various concentration of ChD-3501 in the nematic host E7, was carried out by a UV lamp with wavelength $\lambda_{max}$ = 365 nm and total power 6 W. To study reversible *cis-trans* photoisomerisation of ChD-3501 in both the ethanol and the nematic host E7 the samples were irradiated with He-Ne laser ($\lambda$ = 633 nm) at 7 W.

The thermal phase diagrams of the solid crystals ChD-3501, ethanol solution and cholesteric mixtures containing various concentrations of the ChD-3501, were studied during the both the cooling and the heating process of samples by using polarizing optical microscopy (POM). Samples were placed in a thermostable heater, based on a temperature regulator MikRa 603 (LLD, 'MikRa', Kyiv, Ukraine). The temperature measurement accuracy was ± 0.1 ºC/min. Temperature was measured with a platinum resistance thermometer Pt1000 (PJSC 'TERA', Chernihiv, Ukraine). The speed of temperature change was 0.01 ºC/min (0.1 ºC per 10 min).

Differential scanning calorimetry (DSC) studies were performed using a Mettler DSC 1 microcalorimeter (Mettler Toledo, Switzerland). The cholesteric mixtures (~ 20 mg) were placed into an aluminum crucible, sealed, and the thermograms were recorded in consecutive scans on heating and cooling (scanning rate 5 °C/min). The procedure was repeated 4-5 times. Experimental error was ± 0.1 ºC for the isotropic transition temperature and ± 0.2 ºC for the transition to CLC phase.

To measure the contact angles *β* of various cholesteric mixtures with different concentration of the ChD-3501, a simple method was used. [103] It is based on the measurement of linear dimensions (namely, diameter *D* and height *H*) of the droplet placed upon the dry quartz plate using a horizontal microscope.

To record the transmission spectra of ethanol solution of ChD-3501 and induced cholesteric mixtures at a constant temperature, UV-VIS spectrophotometer (home-made at the Institute of Physics, Kyiv, Ukraine) was used. The spectra were recorded as a function of wavelength over a range of 250 - 800 nm.

## 3. Results and discussions

### 3.1. Synthesis of light-sensitive ChD-3501

In order to synthesize 2-(1R,2S,5R)-[(2-isopropyl-5-methylcyclohexyl)oxy]-2-oxoethyl 4-{(E)-[4-(decyloxy)phenyl]diazenyl}benzoate (ChD-3501), three main steps were carried out. ChD-3501 has been synthesized starting from 4-{(E)-[4-(decyloxy)phenyl]diazenyl}benzoic acid (compound **1**) as shown in Figure1. This azobenzene fragment was synthesized as described in [95] in more detail. Owing to the azobenzene-fragment, the molecule will be expected to be light-sensitive, undergoing the reversible *trans-cis* isomerization.

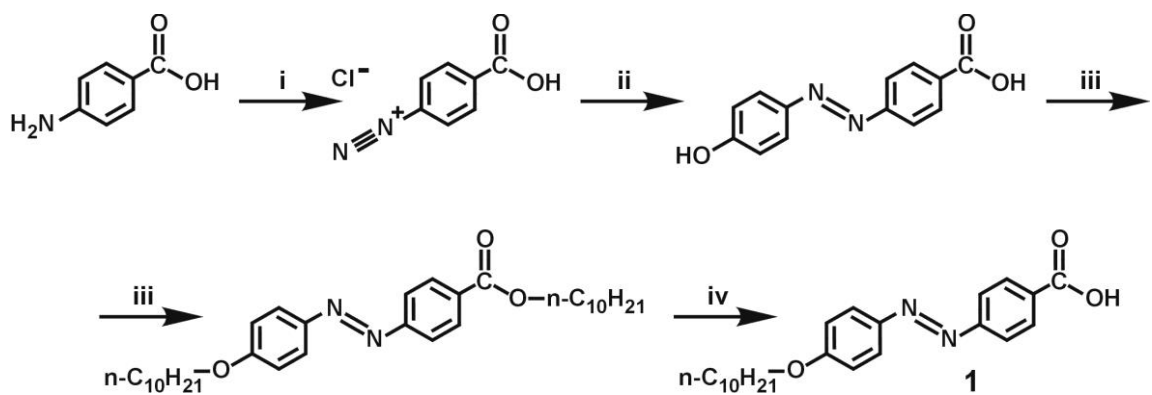

Figure 1. Synthesis of 4-{(E)-[4-(decyloxy)phenyl]diazenyl}benzoic acid (**1**).

Chiral *l*-menthyl chloroacetate (compound **2**) was synthesized by using the method described in detail [13]. For this purpose the reaction between *l*-menthol and chloroacetyl chloride was carried out (Figure 2).

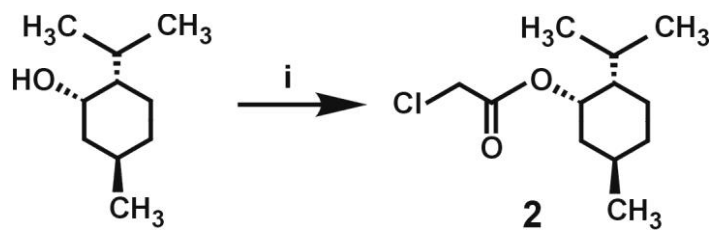

Figure 2. Synthesis of *l*-menthyl chloroacetate (**2**).

It should be recalled that *l*-menthol possesses high chirality as shown in, [6] and it was also used to provide the high chirality of the final molecule. In addition, the chloroacetyl chloride has been used in order to ensure the conformational flexibility and, obviously, to increase of chirality of the final molecule.

Finally, the synthesis of the 2-(1R,2S,5R)-[(2-isopropyl-5-methylcyclohexyl)oxy]-2-oxoethyl 4-{(E)-[4-(decyloxy)phenyl]diazenyl}benzoate (compound **3**) was carried out as shown in Figure 3.

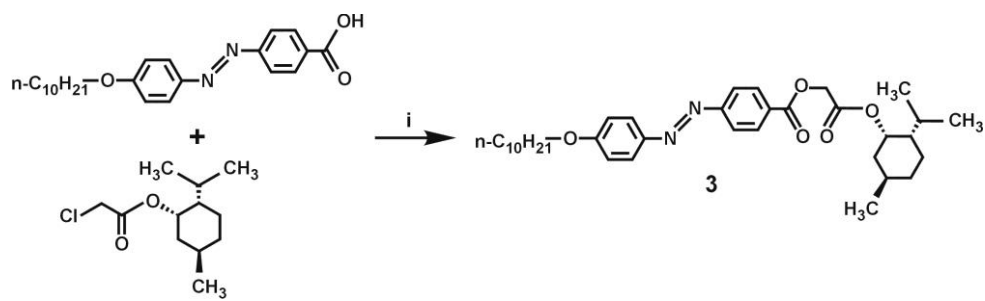

Figure 3. Synthesis of 2-(1R,2S,5R)-[(2-isopropyl-5-methylcyclohexyl)oxy]-2-oxoethyl 4-{(E)-[4-(decyloxy)phenyl]diazenyl}benzoate (**3**).

For this purpose the mixture of 0.38 g (1 mmol) of 4-{(E)-[4-(decyloxy)phenyl]diazenyl}benzoic acid (compound **1**), 0.23 g (1 mmol) *l*-menthyl chloroacetate (compound **2**), 0.41 g (3 mmol) potassium carbonate in 30 mL of the dry DMF was stirred at room temperature during 12 hours and then the mixture was diluted with 100 mL water and extracted with methylene chloride (3×30 mL). The combined organic phase was dried over magnesium sulphate. After removal of the solvent in vacuum, the residue was purified by means of column chromatography on silica gel using hexane as an eluent. The small orange-white solid crystals (0.1 – 0.5 mm length) of the ChD-3501 with yield 67% were obtained.

$^1$H-NMR: δ (ppm) 8.18 (d, $^3J_{H-H}$ = 8.0 Hz, 2H), 7.93 (t, $^3J_{H-H}$ = 7.2 Hz, 4H), 7.31 (d, $^3J_{H-H}$ = 8.8 Hz, 2H), 4.93 (dd, $^2J_{H-H}$ = 16.8 Hz, $^4J_{H-H}$ = 2.0 Hz, 2H), 4.71-4.66 (m, 1H), 4.14–4.08 (m, 2H), 3.19–3.15 (m, 1H), 1.96–1.62 (m, 6H), 1.43–1.26 (m, 16H), 1.07–0.99 (m, 2H), 0.90-0.83 (m, 9H), 0.75 (d, $^3J_{H-H}$ = 8.0 Hz, 3H). IR (cm−1): 3344, 3301, 2984, 2933, 2539, 2484, 2179, 2095, 1696, 1563. The H-NMR spectrum of the ChD-3501 is shown in Figure 4.

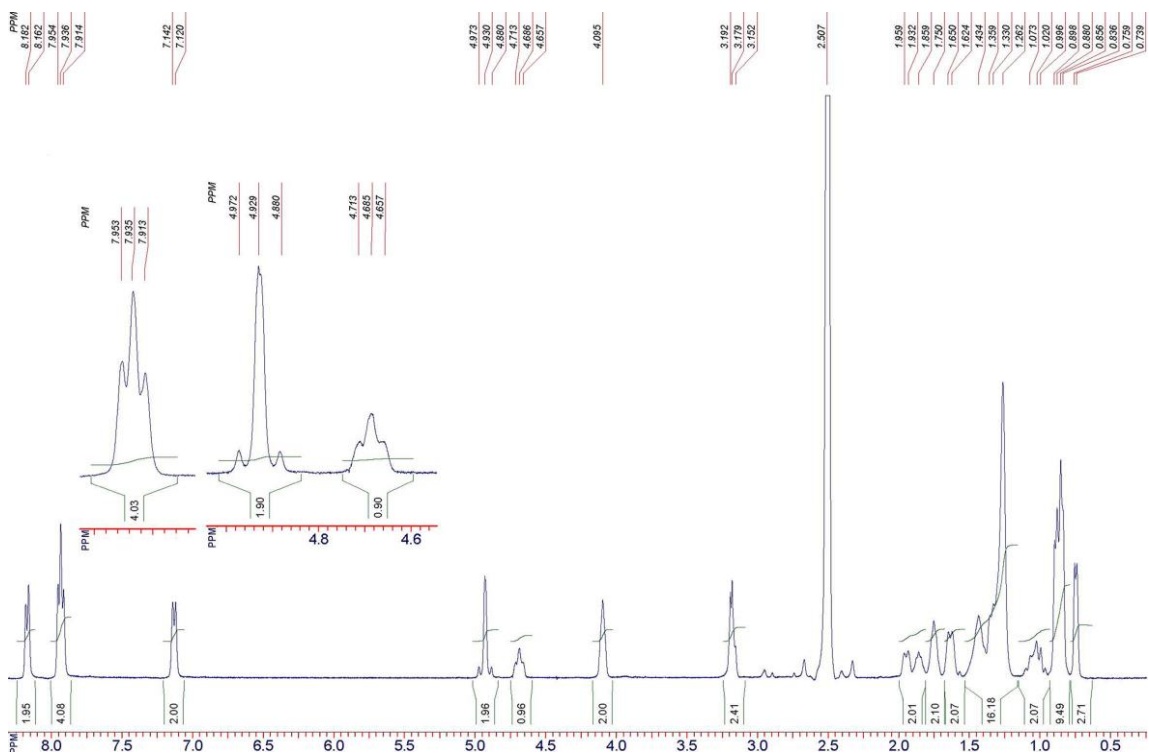

Figure 4. The H-NMR spectrum of the ChD-3501.

*3.2.   UV-VIS irradiation of the ethanol solution of ChD-3501*

In this section the *trans-cis* and *cis-trans* isomerisation of light-sensitive chiral molecules of the ChD-3501 dissolved in ethanol solution under UV/VIS irradiation will be described.

The UV-VIS absorption spectrum ("0" in Figure 5) showed a strong absorption at about 360 nm corresponding to the $\pi-\pi^*$ transition of the *trans*-azo fragment and a weak absorption at 450 nm, which originates from the $n-\pi^*$ transition of the *cis*-azo fragment.

The transformation of absorption spectrum of the ethanol solution of ChD-3501, having the initial OD = 2.3 at the λ = 360 nm under UV lamp irradiation ($\lambda_{max}$ = 365 nm) is shown in Figure 5 (a). Upon UV irradiation, the absorbance at around 360 nm was decreased with an increase in the low-intensity band at around 450 nm and 600 nm due to *trans-cis* isomerization of azo-fragment of the ChD-3501.

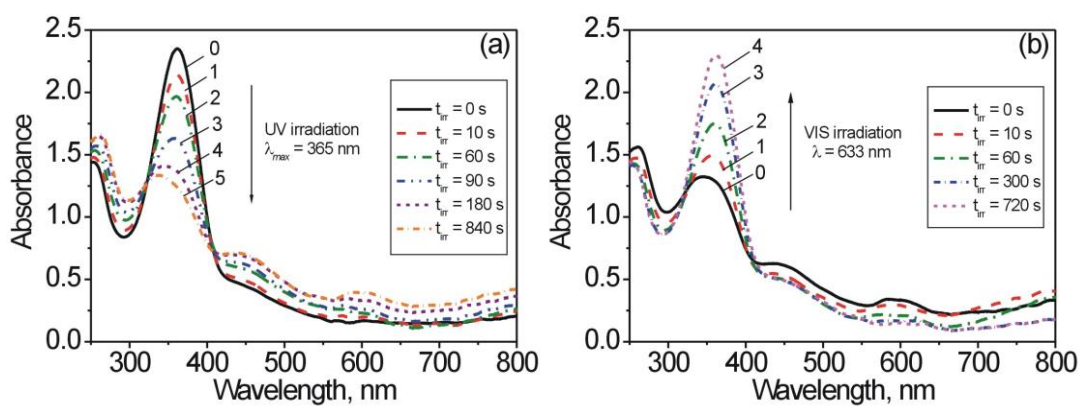

Figure 5. Transformations of absorption spectrum of the ethanol solution of the ChD-3501 upon irradiation by: (a) UV lamp with $\lambda_{max}$ = 365 nm (0 – 0 s, 1 – 10 s, 2 – 60 s, 3 – 90 s, 4 – 180 s, 5 – 840 s) and (b) He-Ne laser with $\lambda$ = 633 nm (0 – 0 s, 1 – 10 s, 2 – 60 s, 3 – 300 s, 4 – 720 s). Thickness of quartz cuvette is 1 cm.

The reason of this decrease in the OD solution is the decrease in the concentration of *trans*-isomer, while the concentration of *cis*-isomer grows. The scheme of *trans-cis* isomerisation of ChD-3105 is shown in Figure 6.

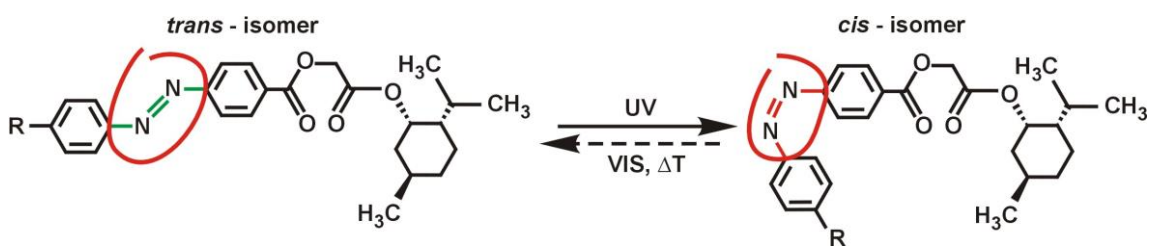

Figure 6. Schema of the *trans-cis* and *cis-trans* isomerisations of the ChD-3501 molecule under UV, VIS irradiation and temperature $\Delta T$.

The reversible *cis-trans* isomerisation (Figure 6) of azo-fragment under VIS irradiation (He-Ne laser, $\lambda$ = 633 nm) leads to an increase in the absorbance of the ethanol solution of ChD-3501 at about 360 nm wavelength range with respect to the initial OD (spectrum 4), as can be seen from Figure 5(b).

*3.3. Temperature influence on the ethanol solution of ChD-3501*

It is known that the *cis*-isomer of the azo-fragment (-N=N-) can undergo the reverse *cis-trans* isomerisation not only due to irradiation by light of higher wavelengths, but also under temperature change $\Delta T$. [16,17,20,104,105]

The reversible *cis-trans* isomerisation of the ChD-3501 molecule under temperature $\Delta T$ is shown schematically in Figure 6. After UV irradiation of the ethanol solution of the light-sensitive molecule ChD-3501, the reversible transformation of absorption spectrum under temperature $\Delta T$ is shown in Figure 7.

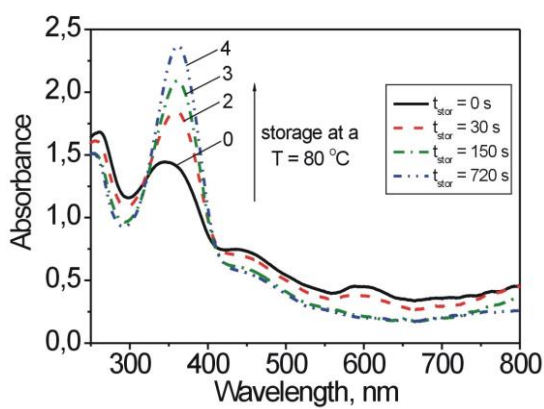

Figure 7. Transformations of absorption spectrum of ethanol solution of the ChD-3501 during storage at a temperature of 80 ºC: 0 – 0 s, 1 – 30 s, 2 – 150 s, 3 – 720 s. Thickness of quartz cuvette is 1 cm. Spectrum 0 was obtained after UV irradiation by lamp with $\lambda_{max}$ = 365 nm and $t_{irr}$ = 840 s.

The initial solution, possessing a high concentration of *trans*-isomer (spectrum 0, Figure 5 (a)), after long exposure $t_{irr}$ = 840 s contains the high concentration of *cis*-isomer of the ChD-3501 (spectrum 5, Figure 5 (a)). Storage at a certain temperature *T* of the ethanol solution of ChD-3501, containing *cis*-isomers, leads to the reversible *cis-trans* isomerisation of the azo-fragment as shown in Figure 6. As can be seen from Figure 7, the reversible photoreaction will be accompanied by the increasing of the OD of solution in the 300 – 400 nm wavelength range. For example, after storage of the ethanol solution of ChD-3501 at *T* = 80 ºC during $t_{stor}$ = 720 s the initial OD = 2.3, which is close to the *trans*-isomer ChD-3501 before UV irradiation (spectrum 4, Figure 7).

*3.4. Induced cholesteric phase based on ChD-3501. Influence of UV radiation and temperature on the cholesteric helix*

It is well known that the dissolution of various chiral dopants in nematic LCs leads to the induction of the cholesteric phase, characterized by the certain pitch of helix *P*, HTP and twist sense. [5,38,39]

The dissolution of ChD-3501 in the nematic host E7 leads to the forming of helical structure, which can be observed under polarising optical microscope (POM) as a "*fingerprint*" texture in a parallel LC cell, possessing substrates with homeotropic (*vertical*) alignment, and Grandjean-Cano (*planar*) texture in a wedge-like LC cell, consisting of pair of substrates with tangential boundary conditions. The "*fingerprint*" and Grandjean-Cano textures of the induced cholesteric mixture, containing 1.8 wt.% ChD-3501 in nematic host E7, are shown in Figure 8.

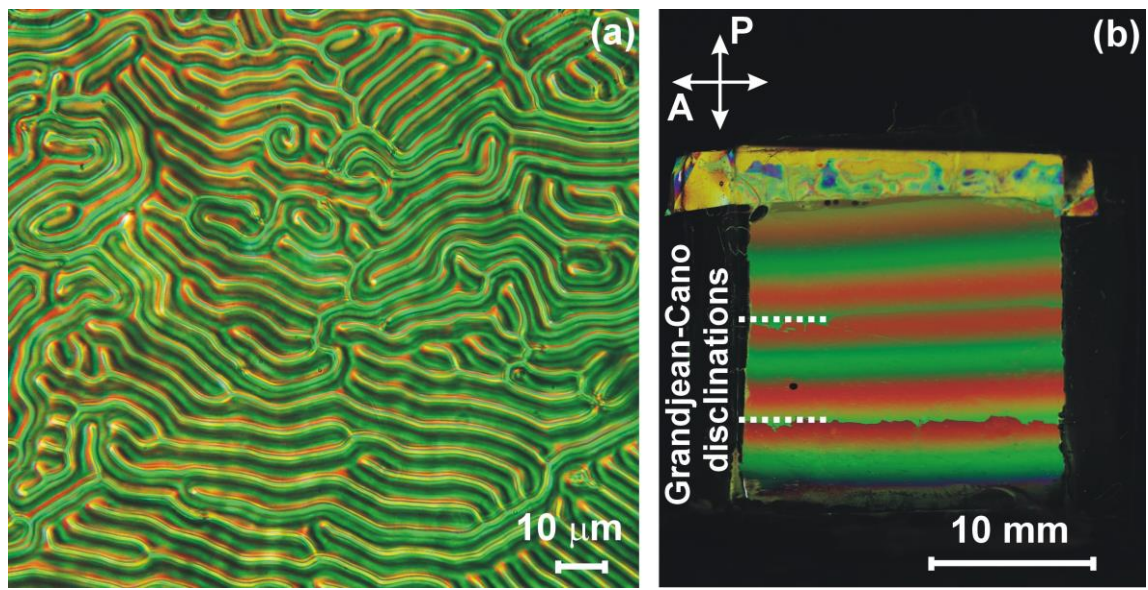

**Figure 8.** Photographs of induced cholesteric mixture based on 1.8 wt.% ChD-3501 and 98.2% E7 under polarizing optical microscope: (a) "*fingerprint*" texture in 13.7 μm parallel LC cell, possessing homeotropic alignment and (b) Grandjean-Cano (or *planar*) texture in wedge-like LC cell ($d_0$ = 5.1 μm and $d$ = 18.7 μm). Textures were observed between crossed polarizers.

The determination of the helix screw sense of cholesteric mixture was carried out by means of the POM. A small prysmoidal solid crystal of ChD-3501 having length about 0.1 mm was gradually dissolved at the top of droplet of the nematic host E7. During the process of dissolution the clockwise rotation of the solid crystal of chiral dopand ChD-3501 in POM was observed. As described elsewhere in detail, [99-101] the clockwise rotation of small crystals of the various ChDs is typical for left-handed cholesteric helix induced by these chiral dopants. In addition, to determine the helix screw sense using the Grandjean-Cano method, [102] the wedge-like LC cell filled by cholesteric was placed between crossed polarizers. In that case the cholesteric mixture based on ChD-3501 and nematic host E7 also displayed the left-handed helix.

The dependence of the cholesteric helical pitch *P* on the concentration of ChD-3501 is shown in Figure 9.

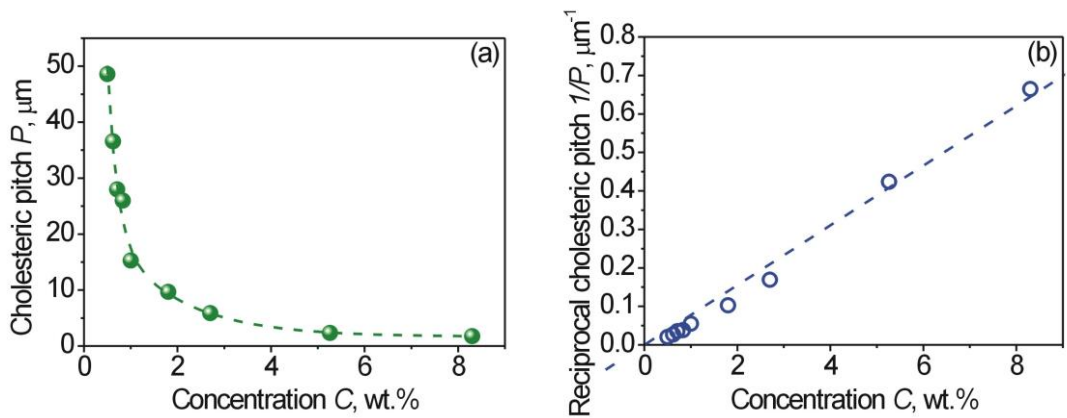

Figure 9. Dependences of the (a) cholesteric pitch *P* and (b) reciprocal cholesteric pitch *1/P* of the induced cholesteric mixtures based on the nematic host E7 on concentration of the ChD-3501.

As can be seen from Figure 9 (a), increased concentration of ChD-3501 in the nematic host E7 leads to the cholesteric helix with shorter pitch *P*. However, the cholesteric pitch *P* at about 8.3 wt.% of the ChD-3501 in the nematic host E7, is not short enough (*i.e. P* ~ 1.7 μm), so that no Bragg reflection in the visible light spectrum is observed. It follows that the helical twisting power (HTP) of ChD-3501 is rather low. However, the light-sensitive ChD-3501, possessing the reversible photoisomerisation can be useful to apply in Raman-Nath type of diffraction gratings with controllable period of grating *Λ* by means of electric field [42-44,58,59] or/and radiation, as was demonstrated for the cholesteric mixture, based on of the chiral dopant PBM and the nematic E7. [30,36]

It should also be noted that the solubility of chiral dopant ChD-3501 is rather low. The fact of low solubility of ChD-3501 is confirmed by more widened DSC peak shapes as will be shown below. In addition, the wettability of induced cholesteric mixtures with ChD-3501 concentrations above ~ 8 wt % is weak, causing non-uniform filling of the LC cells. The dependence of the contact angle of cholesteric droplet placed on quartz plate upon the concentration of dopant ChD-3501 will be shown in Figure 12.

In Figure 9 (b) the linear dependence of the reciprocal cholesteric pitch $1/P$ passing through the origin of coordinates is shown. It is known that the tangent of tilt angle of this linear dependence is the value of HTP ($\beta$) of ChD dissolved in the nematic host. [39] For the ChD-3501 dissolved in nematic E7 the average value of HTP is about – 3.28 (μm × wt. %)$^{-1}$.

Since the ChD-3501 is light-sensitive to UV and VIS radiation (Figure 5), therefore the influence of various spectral range of light on the length of cholesteric helix $P$ was studied.

The influence of the UV irradiation with $\lambda_{max}$ = 365 nm on the pitch of cholesteric helix $P$ of the induced cholesteric mixture based on 1 wt.% ChD-3501 and 99 wt.% of the nematic host E7, is shown in Figure 10(a). During UV irradiation the *trans-cis* isomerization of ChD-3501 occurs (Figure 6), which leads to the unwinding of the cholesteric helix.

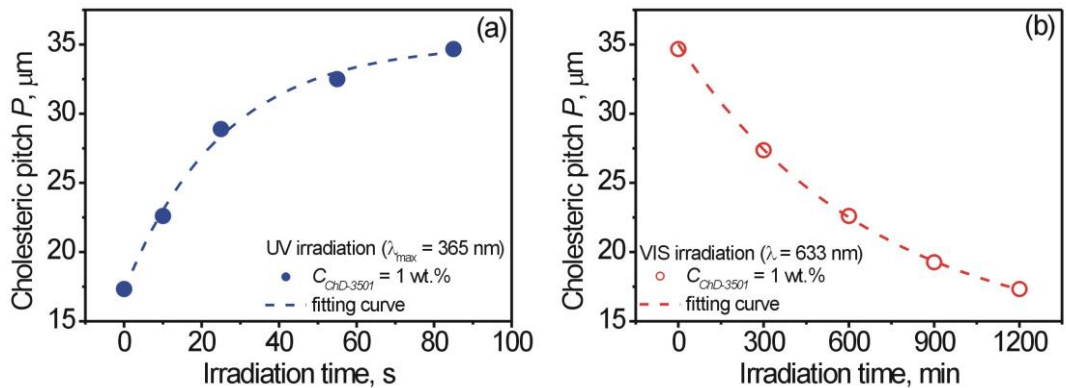

Figure 10. Dependence of the length of cholesteric pitch $P$ on irradiation time: (a) UV-lamp with $\lambda_{max}$ = 365 nm (solid blue circles); (b) He-Ne laser $\lambda$ = 633 nm (open red circles).

By VIS irradiation (He-Ne laser with $\lambda = 633$ nm), the reversible *trans-cis* isomerization of ChD-3501 is realized (the chemical mechanism is shown in Figure 6). In this case the winding of cholesteric helix occurs, and the cholesteric pitch *P* is decreasing (Figure 10 (b)). To restore the initial value of the pitch, longer irradiation time is required in the reverse process. The irradiation time can depend on the radiation source power, the shielding effect during the process of photoisomerisation, various quantum yields of photoreactions, *etc*.

In addition, the *cis-trans* isomerisation of ChD-3501 can happen under the influence of a temperature. In Figure 11 (a) the dependence of the length of helix pitch *P* of the cholesteric mixture, containing $C = 1$ wt.% of the ChD-3501, on storage time $t_{stror}$ of LC cell at $T = 55$ °C is shown. The unwinding and winding processes of the cholesteric helix are cyclic under the influence of the both the light of various wavelengths and the temperature. The cyclic processes of the unwinding under UV irradiation with $\lambda_{max} = 365$ nm (opened blue circles) and winding of the cholesteric helix under the influence of the temperature 80 °C (solid red squares) is shown in Figure 11 (b).

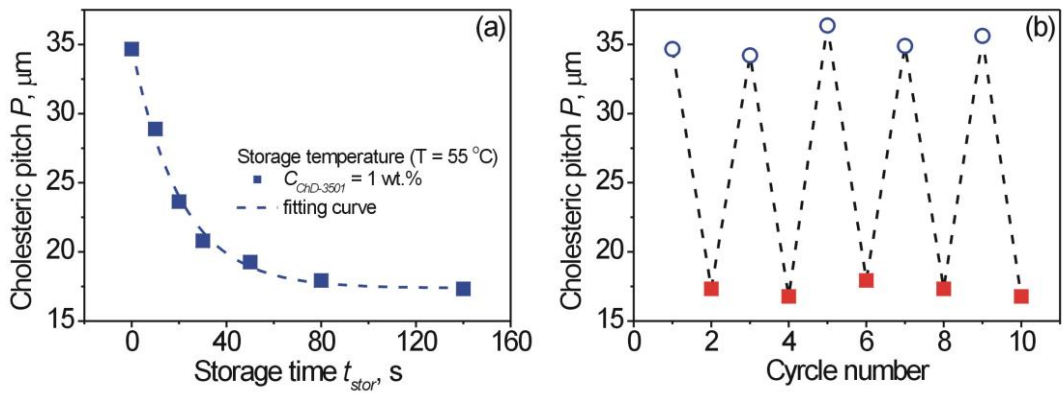

Figure 11. (a) Dependence of the length of cholesteric pitch *P* on storage time $t_{stor}$ at $T = 55$ °C. (b) The cyclic process of the unwinding (opened blue circles) and winding (solid

red squares) of cholesteric helix of the mixture based on 1 wt.% ChD-3501 adding to the nematic host E7.

*3.5 DSC studies of induced cholesteric mixtures*

In this section, the phase transition temperatures between Iso and CLC of the induced cholesteric mixtures, consisting of the nematic host E7 doped by various concentrations of the ChD-3501, were determined by means of DSC and POM.

It is known that the adding of ChD to a nematic host can result in the decrease of the temperature of phase transition [5,38,39], due to reducing of the order parameter of system. These characteristics of the mixture are also functions of concentration *C*.

DSC thermograms of the nematic host E7 and induced cholesteric mixtures, containing various concentrations of the ChD-3501, are shown in Figure 12 (a). The effect of ChD-3501 concentration on the CLC → Iso transition temperature $T_{Iso}$ is observed. Adding of some amount of the ChD-3501 to nematic host E7 leads to a shift of the DSC peak compared with pure E7 (*i.e.* ChD-3501 concentration *C* = 0 wt.%) as shown in Figure 12 (a). This results from the decreasing of the order parameter of nematic host E7 after dissolution of some amount of the ChD-3501, resulting in the decreasing phase transition temperatures $T_{Iso}$ of cholesteric mixture, as can be seen from Figure 12 (b). Temperature values $T_{Iso}$ measured by means of the DSC (solid black spheres) are in agreement with $T_{Iso}$ values obtained from POM (opened blue squares, Figure 12 (b)). The hysteresis of the $T_{Iso}$ on heating and on cooling of cholesteric mixtures was found for both the DSC and POM studies. As an example, for the POM studies hysteresis of the $T_{Iso}$ under cooling (opened blue squares) and heating (opened red circles) processes are shown in Figure 12 (b).

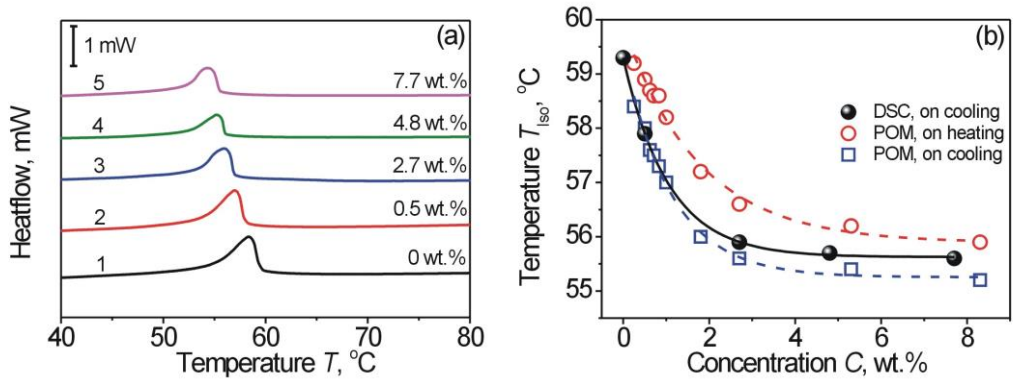

Figure 12. (a) DSC thermograms during the cooling process of the induced cholesteric mixtures, consisting of nematic host E7 doped by various concentrations $C$ of the ChD-3501: (1) - 0 wt.%, (2) – 0.5 wt.%, (3) – 2.7 wt.%, (4) – 4.8 wt.% and (5) – 7.7 wt.%. (b) Dependences of the changes in phase transition temperatures between isotropic (Iso) and nematic/cholesteric (N/CLC) phases in the induced CLC mixture on concentration $C$ of the ChD-3501 obtained by DSC studies (solid black spheres) and POM observations on heating (opened red circles) and on cooling (opened blue squares). The curves are guide to the eye.

*3.5 Study of wettability of the induced cholesteric based on ChD-3501*

In this section we consider the effects of ChD-3501 concentration of the wettability properties of the induced cholesteric. We measured contact angles of the cholesteric mixture droplets with various concentration of ChD-3501 placed on the quartz plate.

As mentioned above, the solubility of ChD-3501 is limited. At concentrations not exceeding about 8 wt.% of ChD-3501, the induced cholesterics can form uniform solutions that are rather stable in time. Higher ChD-3501 concentration leads to non-stable and non-uniform mixtures due to crystallization of the chiral dopant on the one hand and to non-uniform filling of LC cell because of low wettability of the induced cholesteric mixture. In [106] the critical surface tension was determined for thin layers of cholesterol esters forming liquid crystals. It was found that cholesterol esters showed moderately hydrophobic properties. It was interesting to study the contact angle of the droplet of various cholesteric mixtures with different concentration of ChD-3501. The calculation of the contact angle was carried out by measuring the linear dimensions (*i.e.* diameter *D* and height *H*) of a droplet placed onto the quartz plate, as described in detail. [103]

The concentration dependence of the contact angle *α* of the cholesteric droplets containing various amounts of the ChD-3501 is shown in Figure 13.

It should be noted that no effects of irradiation on the value of the contact angle *α* of the cholesteric droplets have been found.

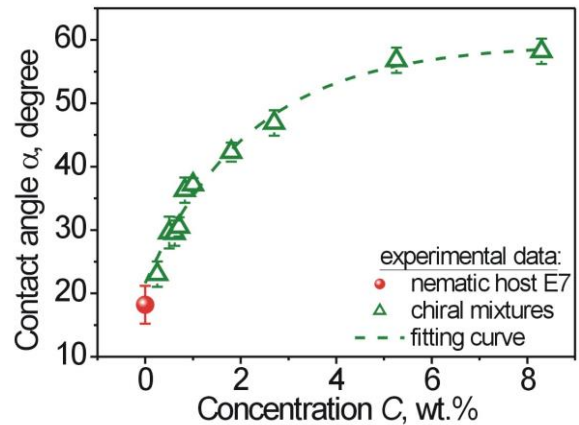

Figure 13. Dependence of the contact angle *α* of the cholesteric droplets placed on the quartz plate without any boundary conditions on the concentration of chiral dopant ChD-3501.

*4.7 Undulation structures in cholesteric mixtures in alternating electric field*

In this section the undulation of cholesteric helix in alternating electrical field will be described. It well known that 1D or 2D undulations of cholesteric helix are observed due to certain ratio *d/P* between thickness *d* of plane-parallel LC cell and the cholesteric pitch *P*. [56]

The period of undulations $\Lambda$ in our cholesteric mixtures as function of ChD-3501 concentration is shown in Figure 14 (a). For all concentrations, the ratio $d/P$ was about 0.86 - 0.9. The frequency $f$ of alternating electric field was 1 kHz, while the value of voltage $U$ when the undulations appeared was varied. As can be seen from Figure 14 (a), higher concentration leads to shorter undulation period $\Lambda$. The minimum value of the undulations period $\Lambda = 30$ µm was observed at $C = 8.3$ wt.% of ChD-3501.

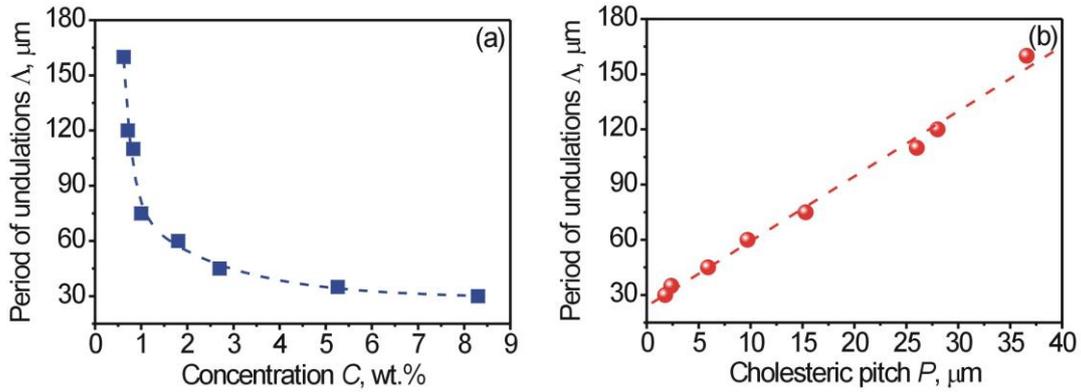

Figure 14. (a) Dependence of the period of undulations $\Lambda$ on concentration of ChD-3501 in cholesteric mixtures based on nematic host E7. (b) Dependence of the period $\Lambda$ of undulations on cholesteric pitch $P$.

In addition there is a certain correlation between the period of undulations $\Lambda$ and cholesteric pitch $P$ as shown in Figure 14 (b). The linear dependence of $\Lambda$ vs. $P$ was found, as can be seen from Figure 14 (b). This correlation can be useful to obtain the minimum value of the period undulations for a cholesteric mixture possessing a certain cholesteric pitch $P$ with the aim to use it as a cholesteric diffraction grating of the Raman-Nath type.

The photographs of 1D and 2D undulations of the cholesteric mixtures containing a certain concentration of ChD-3501, which were formed cells of thickness $d$ in alternating electric fields with various voltages $U$ are shown in Figure 15.

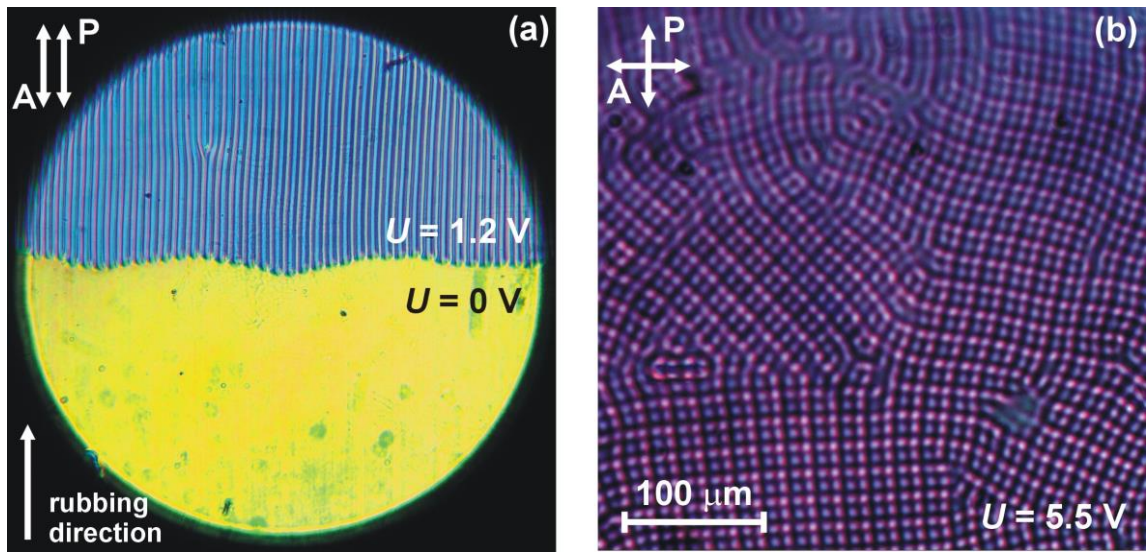

Figure 15. Photographs of the cholesteric mixture based on ChD-3501 and nematic host E7 in alternating electric field with frequency $f = 1$ kHz. (a) 1D undulations structure of the cholesteric mixture with 2.7 wt.% of the ChD-3501 at the voltage of 1.2 V. Cholesteric mixture with $P \sim 5.9$ µm was filled in 7.5 µm LC cell ($d/P \sim 1.3$). (b) 2D undulations of the cholesteric mixture, consisting of 5.3 wt.% ChD-3501, at the voltage of 5.5 V in 20 µm LC cell. The pitch of the cholesteric helix $P = 2.4$ µm and ratio $d/P$ is about 8.3.

As shown recently, the usage of light-sensitive cholesteric mixtures can be of interest for application in the Raman-Nath type of diffraction gratings, which are characterized by the diffraction period $\varLambda$ and various types of grating (*i.e.* 2D, 1D⊥ and 1D∥) switchable by means of both the alternating electric field and the light. [36] Using the light-sensitive ChD-3501 with its reversible photo-transformations to induce the cholesteric helix can allow realization of the electro- and light-switching of undulation structures for the planar texture of cholesterics, as recently mentioned during investigating of the photosensitive cholesteric mixture, based on chiral dopant PBM and nematic host E7. [36]

Thus, it has been shown that due to the usage of light-sensitive CLC, the changing ratio $d/P$ during UV exposure (for the LC cell with a fixed thickness $d$) leads to sequential transitions of 2D, 1D$_\parallel$ and 1D$_\perp$ undulation structure. In addition, due to *trans-cis* photoisomerisation of PBM molecules during UV exposure, the various initial period of grating $\Lambda$, appearing in the alternating electric field, could be obtained.

The change of the undulation structure period under AC with $U = 1.2$ V and $f = 1$ kHz is shown in Figure 15 for the cholesteric mixture containing 2.7 wt.% of ChD-3501 and 97.3 wt.% of the nematic host E7 under the influence of the UV/VIS irradiation. For the cholesteric mixture with $P \sim 5.9$ µm filled into 7.5 µm thick LC cell ($d/P \sim 1.3$), the undulations structure with period $\Lambda \sim 8$ µm and oriented along rubbing direction was formed at $U = 1.2$ V and $f = 1$ kHz (Figure 16 (a)). The illumination of LC cells with undulations structure using various sources of light is schematically shown in Figure 16 (a), (b). By means of UV-lamp (Figure 16 (a)) the initial structure of undulations (Figure 16 (c)) was irradiated. Owing to *trans-cis* photoisomerisation of the ChD-3501 (Figure 6) the light-controllable increase of the initial period of undulations $\Lambda$ is realized, as shown in Figure 16 (d), (e).

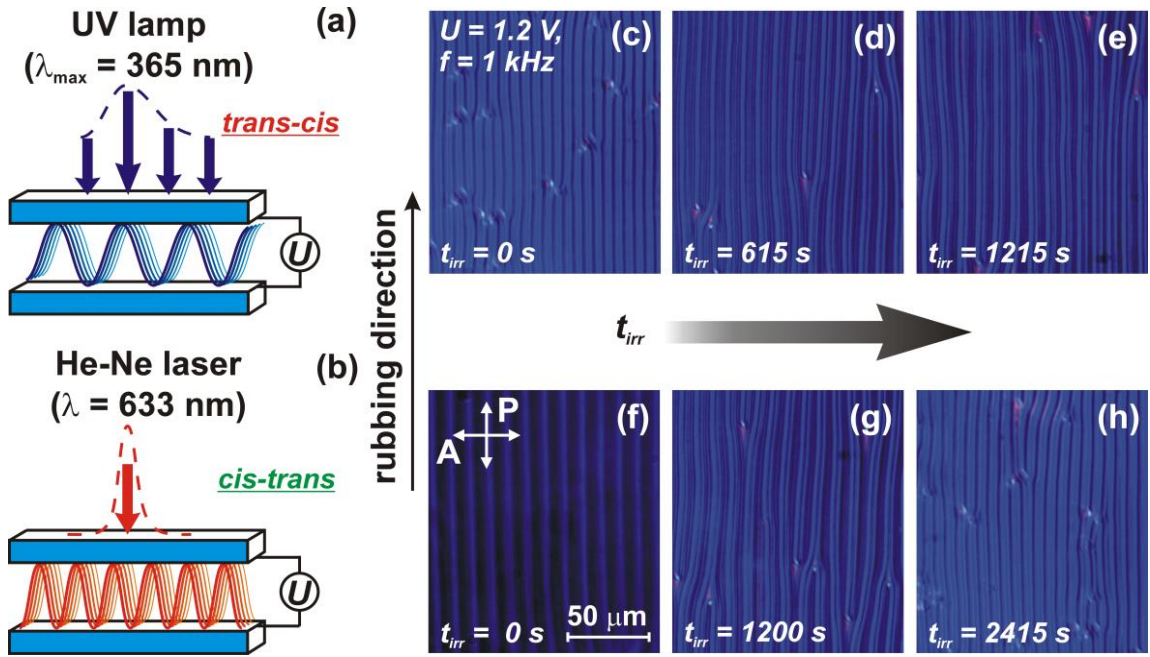

Figure 16. Schema of illumination of the LC cell in AC electric field with $U = 1.2$ V and $f = 1$ kHz by means of: (a) the UV lamp ($\lambda_{max} = 365$ nm) and (b) the He-Ne laser ($\lambda = 633$ nm). Photography of changing of the period of undulation of the LC cell, filled by cholesteric mixture (2.7 wt.% of the ChD-3501 and 97.3 wt.% of the nematic host E7) during: (c)-(e) *trans-cis* isomerisation (UV illumination) and (f)-(h) *cis-trans* isomerisation (VIS illumination).

The dependence of the period of undulations $\Lambda$ on the irradiation time $t_{irr}$ is shown in Figure 17. The monotonous increasing of the undulation period from 8 to 11 µm during the irradiation of sample by UV lamp is shown in Figure 17 (a). The reversible *cis-trans* photoisomerisation of the ChD-3501 due to illumination of sample by He-Ne laser (Figure 17 (b)) leads to sequential decrease of the period of undulation $\Lambda$, as can be seen from Figure 16 (f)-(h). The monotonous decreasing of undulation period to the initial state ($\Lambda \sim 8$ µm) occurs (Figure 17 (b)).

As can be seen from Figure 16 and Figure 17, the reversible light-controlling of the period of Raman-Nath diffraction grating by means of various sources of light can be realized by the usage of the ChD-3501.

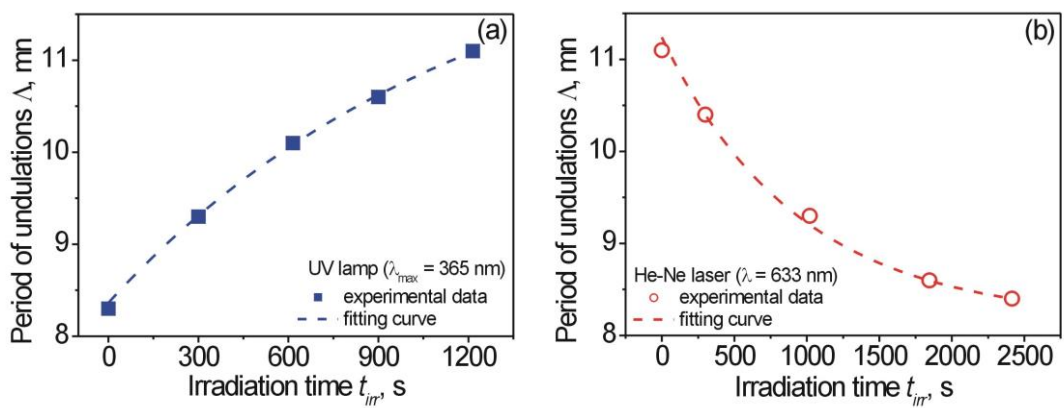

Figure 17. Dependence of the period of undulations $\Lambda$ of cholesteric mixture (2.7 wt.% ChD-3501 and 97.3 wt.% E7) placed in AC electric field with $U = 1.2$ V and $f = 1$ kHz on the irradiation time $t_{irr}$ by: (a) UV lamp (solid blue squares) and (b) the He-Ne laser (opened red circles).

**Conclusion**

In this manuscript we described a newly synthesized 2-[(2-isopropyl-5-methylcyclohexyl)oxy]-2-oxoethyl 4-{(E)-[4-(decyloxy)phenyl]diazenyl}benzoate (ChD-3501), consisting of azo- and aliphatic fragments together with chiral center based on *l*-menthol as the reversible light-controllable chiral dopant. This was the first attempt at combining the chirality of *l*-mentol with *trans-cis* isomerization of the azo-fragment in the development of a photoresponsive dopant. The photoisomerisation of ChD-3501 in ethanol and nematic E7 was studied. The characterization of induced cholesteric mixtures, based on various concentration of the ChD-3510 dissolved in nematic host

E7, was carried out. It was found that the ChD-3501 possesses weak helical twisting power (HTP) $\beta$ = -3.28 (μm × wt. %)$^{-1}$, with no Bragg diffraction in the visible light spectrum. However, we have shown that cholesteric mixtures based on ChD-3501 can be used for various applications, in particular for the creation of Raman-Nath diffraction gratings possessing reversible electro- and light-controllable period as it was previously described for *l*-menthol (PBM) as a non-reversible chiral dopant. [36]

Acknowledgements

The authors thank W. Becker (Merck, Darmstadt, Germany) for his generous gift of nematic liquid crystals E7, A. Kratz (Merck, Germany) for her produce of the Licristal brochure and field service specialist IV V. Danylyuk (Dish LLC, USA) for his gift of some Laboratory equipments. The authors thank Prof. A. Tolmachev (Chemico-Biological Center Taras Shevchenko National University of Kyiv), Dr. S. Minenko (Institute for Scintillation Materials of STC "Institute for Single Crystals" of the NAS of Ukraine) and Dr. S. Lukyanets (Institute of Physics, NAS of Ukraine) for the helpful discussions.